\documentstyle[11pt,newpasp,twoside,epsf]{article}
\def\edcomment#1{\iffalse\marginpar{\raggedright\sl#1\/}\else\relax\fi}
\reversemarginpar

\begin{document}
\title{Ground-Based Optical Deep Pencil Beam Surveys}

\author{P.C. Boeshaar} 
\affil{Drew University}
\author{V. Margoniner}
\affil{Bell Laboratories, Lucent Technologies}
\author{and the Deep Lens Survey Team}

\begin{abstract}

The initial results of the Deep Lens Survey (http://dls.bell-labs.com)
to identify possible brown dwarfs and extremely metal poor red halo
subdwarfs near the hydrogen burning limit are presented. Individual
deep CCD high galactic latitude survey fields appear to offer a low
probability of discovering field BD's, but taken collectively offer an
opportunity to begin addressing questions regarding the scale height
and distribution of these objects.  In all likelihood, the very depth
of such surveys will greatly increase our knowledge of the coolest
extreme halo objects, which currently are known in far fewer numbers
than T dwarfs. Ultimately, the large volume surveyed by the Large
Synoptic Survey Telescope will identify vast numbers of such objects,
providing a more complete picture of their spatial distribution.

\end{abstract}

\section{Introduction}

Since the early 1980's, CCD's have been employed on large telescopes
to obtain deep images of small sections of the sky.  With better than
$60\%$ quantum efficiency ranging from the atmospheric UV cutoff to
the drop in their sensitivity at $1\mu$, these first surveys readily
reached a limiting magnitude of 26 in all of the broad bandpasses
encompassed by this wavelength range.  Until the mid-1990's the 1 cm
size of the CCD chip limited the area on the sky covered with each
exposure to a few sq. arcmin.  Consequently, the actual volume
surveyed per field was very small, typically a few or tens of
$pc^{3}$, up to the scale height of 325 pcs for late type disk
population dwarf stars.  As such, they proved an ineffective method to
discover the latest M dwarfs, let alone L or T brown dwarfs with
unknown scale height and local volume densities estimated to be in
the range of $10^{-2}$ per L or T class of objects (Liu et al. 2002).
Complicating the picture was the fact that these untargeted surveys
had multipurpose science drivers, often were focused at higher
galactic latitudes, and employed many different filter systems.

More recent developments leading to larger CCD chips and mosaics have
increased the sizes of individual deep survey fields to tens of arc
minutes across, but have not significantly increased survey depth.
With the success of the large area SDSS optical survey in discovering
BD's, and the marked absence of similar results from the many deeper
pencil beam surveys over two decades, the latter do not appear to be a
very efficient method to discover such objects.  Yet despite continued
limited spatial coverage, because of their increased depth,
serendipitous discoveries of BD's by existing pencil beam surveys may
help to address questions surrounding the uncertainties regarding the
spatial distribution of field BD's.

\section{Recent Observations}

\begin{figure}
\plotfiddle{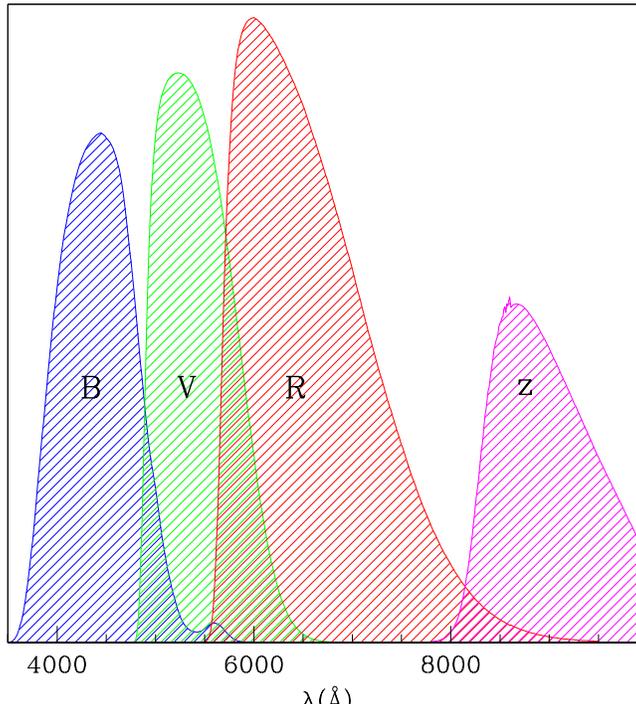}{50mm}{0}{50}{50}{-150}{-200}
\vskip 3.8cm
\caption{DLS filter response convolved with CCD and telescope response
curves.}
\end{figure}

This paper will focus on the initial results of the Deep Lens Survey
(DLS, http://dls.bell-labs.com).  The DLS is composed of 7 ``random''
fields to be surveyed over a five year period, each 2 degrees on a
side, selected to avoid bright stars and areas with high galactic
extinction.  Taken with Mosaic CCD imagers on NOAO's Blanco and Mayall
Telescopes, each field is split into a 3x3 array of roughly 35
sq. arcmin Mosaic-size subfields (0.257 arcsec per pixel) called p11
through p33.  These subfields are imaged in four bands: B, V, R and
z'. See Fig. 1 for the combined filter plus CCD plus telescope
response curves.  Twenty 600 second exposures are obtained in each of
the B, V, and z' filters, with twenty 900 second exposures in the R
filter.  We estimate limiting magnitudes (Vega) for 5 sigma stellar
detection in 1 arcsec seeing of 26.3 in B, 26.4 in V, 26.5 in R and
24.1 in z', with seeing to date averaging 0.8 arcsec in R and 1.0
arcsec in the other filters.

The data reduction has been completed for several subfields and is
available for public access at the DLS web site listed previously.
This site contains information regarding the survey specifics (such as
field coordinates, completeness, etc), matched catalog 2.5 arcsec
aperture magnitudes produced by SExtractor (version 2.1.6) for objects
detected in each released subfield, as well full images in each filter
for download, and the tools for analysis.  One of our survey fields
overlaps one in the NOAO Wide Deep Survey (Tiede et al. 2002) and will
have K-band coverage also.

\subsection{Initial Results in Detection of Brown Dwarfs}

\begin{figure}
\plotfiddle{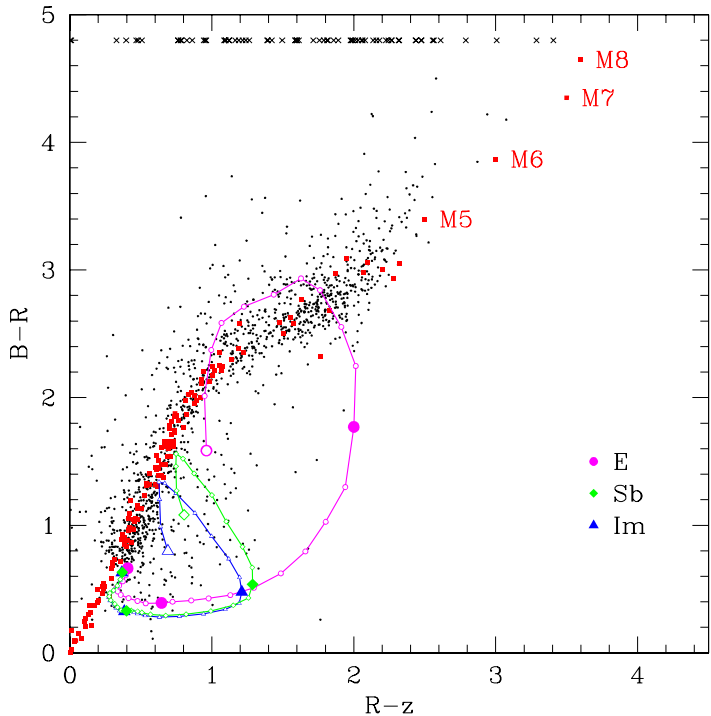}{50mm}{0}{50}{50}{-150}{-190}
\vskip 3.8cm
\caption{$19<R<26$ stellar objects in the F4p22 subfield.  See text
for definition of symbols.}
\end{figure}

We have completed the data reduction and analysis of two 1600 sq.
arcmin subfield areas, F1p22 ($l^{II} = 125^{\circ}$, $b^{II} =
-50^{\circ}$) and F4p22 ($l^{II} = 257^{\circ}$, $b^{II} =
47^{\circ}$) to search for BD candidates as well as red halo subdwarfs
at/near the hydrogen burning limit. Using average absolute magnitudes
for L and T dwarfs obtained from Dahn et al. (2002), we estimate that
we could detect early L dwarfs ($R-z^{\prime} = 3.6$ or greater) out
to a distance of 410 pc, mid L dwarfs ($R-z^{\prime} = 4$ or greater)
out to 140 pc, and T dwarfs ($R-z^{\prime} > 4.5$) out to 90 pcs. By
way of comparison, using the SDSS z* (AB mag) detection limit of 20.3
alone, the SDSS limiting distances fall to roughly $1\over3$ of our
values.  Searching our images for stellar BVR dropouts (i.e. detected
in a z' image alone) should more than double our line of sight
distance.

Fig. 2 presents the $B-R$ vs. $R-z^{\prime}$ data for all $19<R<26$
objects in the F4p22 subfield morphologically classified as stellar
using an algorithm based on the out-of-roundness deviations.  Similar
results are obtained for F1p22.  Those objects which are too faint for
detection in the bluer band have been indicated by a small $\times$ at
their given $R-z^{\prime}$ color along the top of the plot.  The
filled large squares represent Gunn-Stryker spectrophotometric
standards, with additional notations to indicate the approximate
positions of M5, M6, M7, and M8 dwarfs.  The tracks reflect expected
evolution and K-corrections for elliptical (circle), irregular
(triangle), and Sb+Sc (diamond) galaxies with increasing redshift
(steps of $z=1$ are indicated by filled symbols, starting with the
large open symbol at $z=0$) based on spectral energy distribution
templates from Bruzual and Charlot (1993).  These are shown to
indicate regions free of contamination by compact high redshift
galaxies, a point which will be especially important in searching for
red halo subdwarfs near the hydrogen burning limit. Note that QSO's of
increasing redshift fall well blueward of the stellar locus.

We discovered no objects with colors indicative of BD's, i.e.
$R-z^{\prime}>3.4$ and $V-R>2.4$. Such objects are probably too faint
to be detected in $V$, let alone $B$. Furthermore, assuming that some
BD's would be too faint in our R band for detection, a check was made
in each field for stellar-like objects which dropped out of the R, V,
and B images.  To an R limit of 26.5 mag, we typically find
$200,000-300,000$ galaxies and 2,000 to 3,000 stars per subfield.  We
detect approximately 7,000 BVR dropout objects in each subfield, only
10 of which are stellar.  Upon closer individual inspection, none of
these appeared to be good BD candidates.  Though the DLS subfields
cover slightly more area, go roughly as deep, and had similar seeing
to the IfA Deep Survey subfields (Liu et al. 2002), we detect no brown
dwarf candidates; while they discovered one T dwarf in their first
field.  At a $z^{\prime}=20.9$ mag and $R\sim25.5$, such an object
would have been easily detectable by the DLS.  No additional
detections have been reported (Liu 2002b) even though both surveys
cover a similar range of galactic latitude.

In a similar survey covering a much smaller area, comparable results
for late M dwarfs were found by Boeshaar and Tyson (1986).  No stars
with $R-z^{\prime}>2.8$ (equivalent to $Mv > 15$) were discovered in
twelve 12 sq. arcmin deep CCD fields.  Yet in one high latitude 40
sq. arcmin field Boeshaar et al. (1994) detected six M dwarfs ranging
from dM6 to dM8-9. Furthermore, the late dM's all appear to be located
within an area of less than 20 sq. arcmin.  Few surveys report null
results when searching for objects with low probability of
detection. Often only the serendipitous discoveries are announced. Our
cursory evaluation of late dM, L, T stars found in similar deep CCD
surveys lends support to the contention that small area high galactic
latitude pencil beam fields have a low probability of discovering
large numbers of stars at or below the tail end of the main sequence.

\subsection{Detection of Cool Halo M Subdwarfs}

\begin{figure}

\plotfiddle{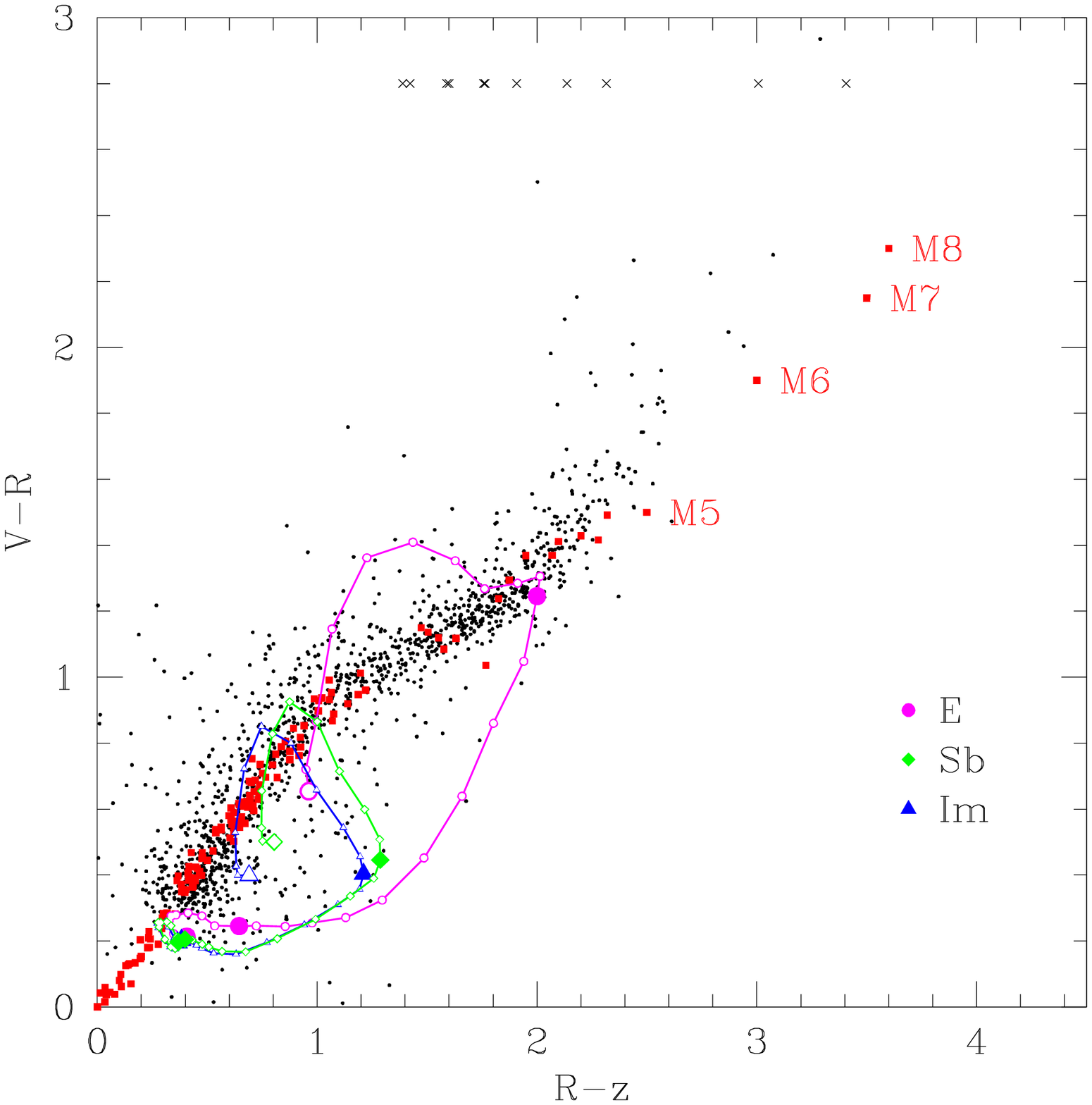}{40mm}{0}{45}{45}{-150}{-170}
\vskip 3.8cm
\caption{Plot of $V-R$ vs. $R-z^{\prime}$ for F4p22 with symbols
defined as in Fig. 2.}

\plotfiddle{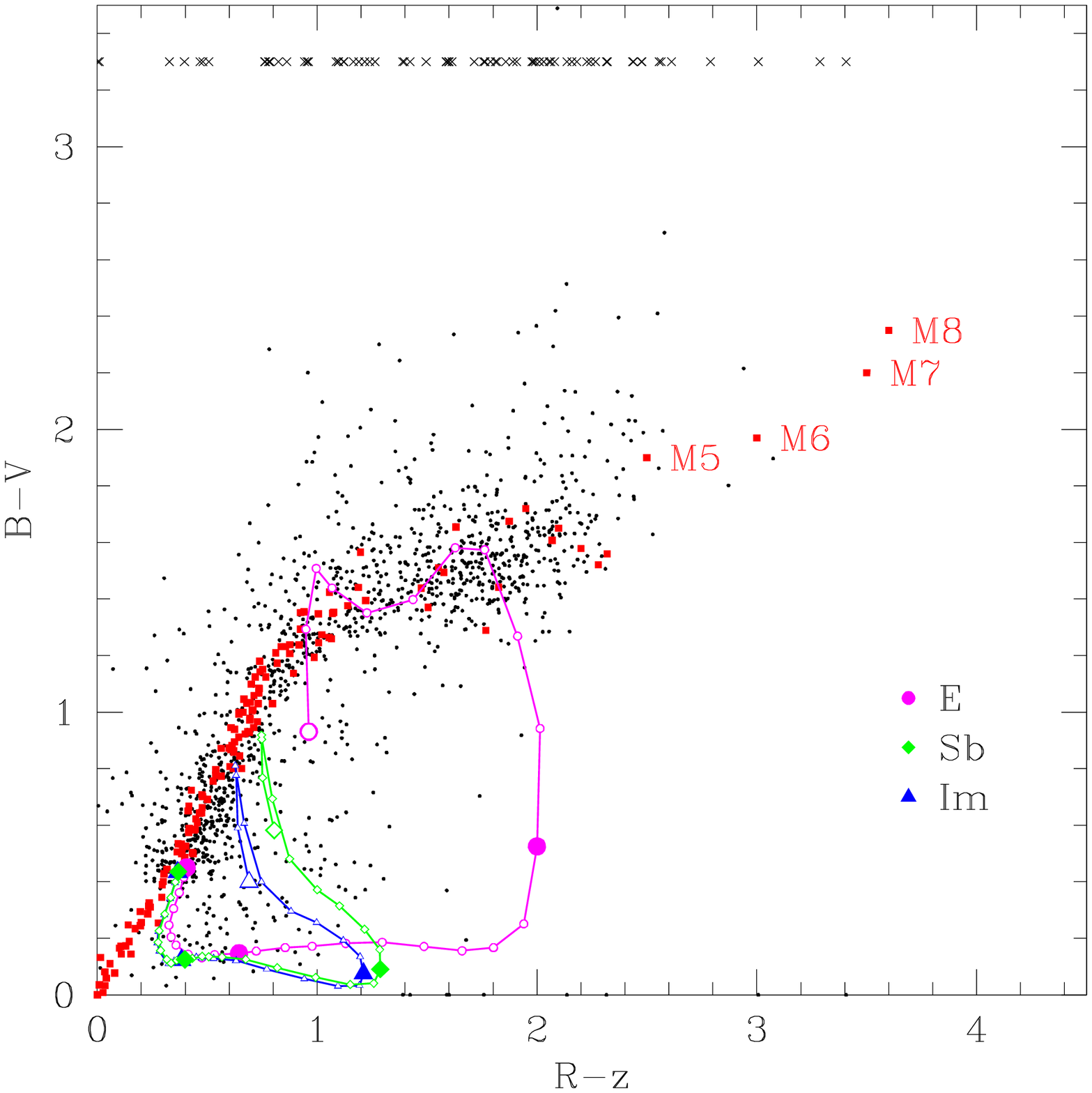}{40mm}{0}{45}{45}{-150}{-200}
\vskip 4.4cm
\caption{Plot of $B-v$ vs. $R-z^{\prime}$ for F4p22 with symbols
defined as in Fig. 2.}

\end{figure}

All extremely red field halo subdwarfs have been identified initially
from photographic high proper motion surveys. To date, there are only
4-5 extremely metal poor subdwarfs with $M_{V}>14$ known (Gizis 1997,
Gizis and Reid 2002). These objects would be the best candidates to
have masses near the halo hydrogen burning limit.  Large uncertainties
remain both theoretically and observationally regarding the relations
between mass, metallicity, and color for these stars.  Consequently,
very large uncertainties in the number density of such objects exist
(Hawley and Gizis 2000). Do red subdwarfs exist at
$[Fe/H]\raisebox{-0.5ex}{$\stackrel{<}{\scriptstyle \sim}$}-2.5$,
comparable to the most metal poor hotter subdwarf population (Beers et
al. 1998)? Answers to these questions will be important to
understanding the early history of low mass star formation in our
Galaxy.  While the initial surveys have sampled only the immediate
solar neighborhood, i.e. out to approximately 100 pc, very deep
multicolor CCD surveys are ideally designed to sample the halo out
beyond 1 kpc for the intrinsically faintest subdwarfs, and beyond 3
kpc for objects of Mv$=12$.

Dahn et al. (1995) have demonstrated that extremely metal poor M
subdwarfs (esdM's) appear well-separated from disk M dwarfs by more
than 0.5 mag in the $B-V$ vs. $V-I$ color plane.  Observing M dwarfs
of known metallicity, Gullixson et al. (1995) show a similar effect
using the broad band $B_j$, $R$ and $I^{\prime}$ bands of a deep
4-meter CCD survey. This population separation is most noticeable
between B and other filters due to the effects of differential
molecular oxide blanketing in the V, R, I and z' filters in comparison
to the B band, which is dominated by atomic lines.  Conversely, the
$B-V$ color is an inaccurate temperature index for M dwarfs.  With
reduced molecular blanketing of the spectra, one expects M subdwarfs
to fall closer to the black body line in broad band color plots.  In
Fig. 2, esdM's of $M_{V}=12-15$ fall in a region from $R-z^{\prime}
\approx 1.5-2.15$, with $B-R>3.3$, well above the $B-R$ colors of the
M disk population.

Since a metallicity gradient exists among the M dwarfs, extending
from young disk to extreme halo (Hartwick et al. 1984), we would
expect to see a continuous distribution of stars in $B-V$ or $B-R$ for
a given red color index, as shown.  Note in Fig. 3, which plots $V-R$
vs. $R-z^{\prime}$ for the same F4p22 subfield, that without B
magnitudes it is more difficult to distinguish among the M main
sequence ($R-z^{\prime}>1$) population classes than in Fig. 1.
Here, changes in molecular blanketing with decreasing metallicity
affect both V and R almost equally at a given temperature. In the
$B-V$ vs. $R-z^{\prime}$ plot of Fig. 4 for our F4p22 subfield, the
separation among the M dwarf distribution becomes more pronounced as
expected.

To address the question of contamination of the ground-based colors
used to identify red subdwarfs by an unresolved population of high red
shift elliptical galaxies at faint magnitudes (Reid et al. 1996), we have
superimposed the redshift-evolutionary plots of several galaxy types
on all of our color-color plots.  The elliptical galaxies fall well
below the esdM regime in Fig 1; and they fall among the disk
population M dwarfs in Fig.3.  Thus negligible galaxy contamination
should be encountered in the region where the extreme M subdwarfs are
located.  In a previous deep CCD survey, Tyson and Boeshaar (1997)
found that the six identified subdwarf candidates lay within the
numbers expected, extrapolating from the luminosity function of Dahn
et al. (1995, 1996).  Follow-up multi-slit spectroscopy on three of
these candidates using the CTIO 4-meter confirmed their identity as M
subdwarfs.  The combination of accurate predictions of galaxy colors,
B filter data, plus careful analysis of the point spread function in
our color images indicates that we should be able to identify M-type
halo subdwarfs with $90\%$ confidence even at our faint magnitude
limits.

\section{Discussion} 

Very deep multicolor pencil beam surveys provide a rich data base
which can be utilized to provide information on the scale height
distribution of field brown dwarfs and more accurately define the
luminosity function of lowest mass stellar halo population.  Though
each individual mosaiced CCD subfield in itself offers a low
probability of detection, the combined data from larger abutted areas,
such as those covering 4 sq. deg. in the DLS, should allow for a
meaningful sample of objects to be evaluated. Using $\sim0.01/pc^3$
each for the local volume density of L and T dwarfs as estimated by
Liu (2000), we might expect to find
$\raisebox{-0.5ex}{$\stackrel{<}{\scriptstyle \sim}$}3$ T dwarfs and
$\raisebox{-0.5ex}{$\stackrel{<}{\scriptstyle \sim}$}33$ L dwarfs in
each 4 sq. degree DLS field when our survey in completed.  This
assumes detection in $R-z^{\prime}$ and a uniform spatial distribution
with a vertical scale height of no more than 200 pcs.  Reducing the
scale height to 100 pcs lowers the expected number of L dwarfs by a
factor of 8. If we simply search for stellar objects which drop out of
all bands except z', then the number of T dwarfs per field should drop
from $\sim30$ (indicating a scale height $\sim200$ pc) to about 4
(scale height $\sim100$ pcs).  Whereas only a relative handful of the
coolest halo subdwarfs have been identified to date using other
techniques, carefully analyzed deep CCD observations including B band
data should discover in one DLS field numbers of very red esdM's
similar to those currently known.  Extrapolating from the luminosity
function of Dahn et al. (1995, 1996), we expect to find at least 4
extreme subdwarfs similar to LHS 1742a (esdM 5.5) or LHS 1826 (esdM6)
per each LHS field and 15 stars similar to LHS 453 (esdM3.5).  The
observations for esdM's should prove equally important as the search
for additional L and T dwarfs since we know far less about the coolest
halo population.

When the 30 sq. degree DLS is finished, more will be known about these
subluminous halo objects and field BD's. The problem with current
pencil beam surveys is that the numbers of halo stars found do not
permit cutting the sample in multiple ways, testing their distribution
with galactic coordinates, etc.  Enabled by a confluence of optics and
microelectronics advances, the proposed Large Synoptic Survey
Telescope ($LSST$, http://lsst.org), will produce a large volume
optical multi-color survey of $14,000$ sq. degrees to $27^{th}$
magnitude.
Completion of the facility is expected by 2012, and the data will be
public.  Image quality is expected to be as good or better than the
current generation of new technology telescopes ($0.4^{\prime\prime}$
fwhm), and each field of the sky will be covered hundreds of times,
permitting proper motion studies and enhanced rejection of galaxies
due to the low surface brightness achieved.  Moreover, due to higher
quantum efficiency at $\sim1\mu$, the $LSST$ filter set will most
likely include an additional filter near this spectral region,
complementing the SDSS $ugriz$ filter set.  The combination of broad
area coverage, survey depth and increased near IR sensitivity will
allow for a direct determination of the local disk distribution of
field brown dwarfs.  Furthermore, the $LSST$ will create a sample of
thousands of subluminous halo stars, probing deep into the halo as
well as spanning the full range of galactic coordinates. The role of
these stars in the evolution of the structure of our galaxy will then
be more apparent.

\acknowledgments

The DLS data are taken on the 4-meter Blanco and Mayall telescopes,
operated by the National Optical Astronomy Observatories (NOAO).  NOAO
is operated by the Association of Universities for Research in
Astronomy (AURA), Inc., under Cooperative Agreement with the National
Science Foundation.

\end{document}